# RELIGION IN THE COSMOLOGICAL IDEAS IN UKRAINE (FROM XI TO XVII CENTURY)

*Koltachykhina Oksana Yu., G.M.Dobrov Center for Scientific and Technological Potential and Science History Studies of National Academy of Sciences of Ukraine*

**Abstract:**
Cosmology, whose object is the Universe, is in contact with the religion closer than any other science. We will try to trace the historical change of views on the universe in Ukraine from XI to XVII centuries and show the influence of religion in the construction of the models of the Universe in that time.
In the early XIth century, a great authority, and the spread had a Byzantine texts. In Ukrainian chronicles were described the structure of the world. The astronomical interpretation of cosmological ideas, a system of Ptolemy, was stated in treatise 'Izbornik' (in 1073 and 1076). In the XV century in Ukraine were spread 'Cosmography' De-Sakrabosko and 'Shestokryl' Immanuel bar-Jakob.
The first course of Nature Philosophy that was read in Ukraine and studied the system of Copernicus was Gizel's philosophy course 'The work of the whole philosophy' (1645-1647).
Th. Prokopovich was also the first who gave an explanation of Copernicus's theory (early XVIIIth century). Despite the fact that in his course taught various systems of the world, but he believed that the world had been created by God. Until the end of the XVIIIth century a key role in explaining the origin and structure of the Universe had a religion and the Bible text.

Cosmology, whose object is the Universe, is in contact with the religion closer than any other science. We will try to trace the historical change of views on the universe in Ukraine from XI to XVII centuries and show the influence of religion in the construction of the models of the Universe in that time.

Formation and development of outlook in Ukraine has a long history and dated the days when people could not actively influence on the nature. People were an integral part of it and completely dependent on it. The biblical texts were of highest importance, but the Holy Scripture did not give answers to specific issues concerning the universe, the properties of things and the origin of phenomena. The sources of information were the works of ancient authors, especially Aristotle. Byzantine culture (IX-X century) made great contribution to systematization of the knowledge available in ancient literature. There was no clean-cut division into specific scientific disciplines. The ancient knowledge is systematized during this period.

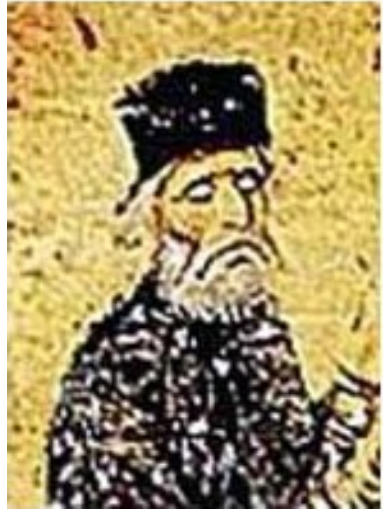

**1 Michael Psellus**

Michael Psellus (1018-1096) was one of the persons who started to criticise the Christian canons. He had the honorary as Consul of the Philosophers (Hypatos ton Philosophon). M. Psellus did not share the ideas of Aristotle and espoused Neoplatonism system of philosophy stating about the Uniqueness, the World Mind and World Soul. He relied on Ptolemy's concept of the geocentric universe, imagining the universe as a set of celestial spheres rotating around the spherical Earth [1–5].

Middle the view of M. Psellus was Simeon Seth. In work 'Overview of Natural beginnings' [5] said he that the Earth rotates around the celestial sphere.

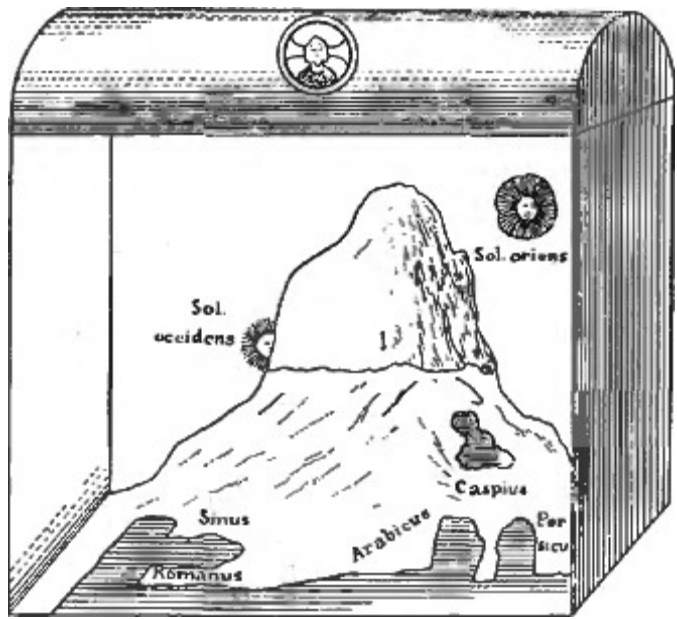

**2 Structure of the World in accordance with the views of Cosmas Indicopleustes [4].**

In Ukrainian chronicles (beginning with the XI-XII) was described the structure of the world. There were several options: 'Christian topography' by Cosmas Indicopleustes (Byzantine cosmography, merchant and later monk), 'Shestodnev' by Joseph Bulgarian Exarch, 'Chronicle' by George Hamartolus [6]. The mentioned works were influenced considerably on the science in Ukraine. The astronomical interpretation of cosmological ideas, a system of Ptolemy, was stated in treatise 'Izbornik'. There were two versions of them (in 1073 and 1076) that were the basic source of knowledge for residents of Kievan Rus (Kyivan Rus'). It is believed that these manuscripts were written in Kyiv, Ukraine. The authors by preparation of these works used as the basis the Byzantine texts. The first part of 'Izbornik' is analyzed astrological concepts. Its second part is included the information about the chronology of various nations and the names of months. In this work is contained the basic provisions of Aristotle's Metaphysics [2].

Cosmas Indicopleustes rejected the geocentric theory of Ptolemy and a statement that 'the Earth has a shape of sphere'. He tried to explain his ideas about the world structure on a basis of the Holy Scripture. C. Indicopleustes believed that the Earth has a shape of a flat rectangle. The Sun is behind a hill. The people live on the slopes of this hill. The Sky is solid and transparent. The Sky has a shape of a tent – it is the First sky. The Second Sky looks like a skin that was stretched on the First Sky. The Sun and the Moon are located below. The Moon shines and does not disappear, but hides its light. The Earth and the Sky are unmoved. The stars, the Sun, the Moon are moving lights. According to Indicopleustes, the World is a two-floor 'building'. The first floor of this building is occupied by the Nature. The Second is separated by the Two-Layer Sky (firmament and water and invisible sky). The Earth looks like the Ark of Noah and the Tabernacle mentioned in the Old Testament. He opposed the ideas of Aristotle and Ptolemy, who believed that the Earth is round [7, p. 8].

Not only were the above-mentioned primitive ideas concerning the structure of Universe but also 'Shestodnev' by Joseph Bulgarian Exarch was public in Ukraine. 'Shestodnev' highlighted the astronomical problems in view of the achievements of antiquity [8, 9]. The first mention about the world appeared in the chronicle of Joseph (XIII century). In 'Shestodnev' is said that the Earth is in the centre of spherical vault, which involves several moving concentric circles. The Circles are attached to the Sun and Moon, five planets ("floating stars") that perform loop-like movements, and unmoved stars. The Sun, Moon, stars and the creation have a shape of ball. The Sun moves along the underground and aboveground floors. Within the year the Sun crosses along 12 zodiacal constellations, moving along 'life-giving' circle. In 'Shestodnev' are contained many astronomical information: data about the size of celestial bodies, explanation such phenomena as equinox and solstice, the change of seasons, deviation in the shadow of the Southern Hemisphere, as well as information related to the climatic zones of the Earth. It contains also information concerning the structure of the world according to Ptolemy and C. Indicopleustes.

Distribution of cosmography tenets is stated in 'Christian topography' by Cosmas Indicopleustes and 'Shestodnev' by Joseph Bulgarian Exarch allow to state that the astronomy in medieval Ukraine had two trends: one trend – primitive explanation of correspondent tenets of the Holy Scripture, the second

one – understanding of astronomical knowledge of the ancient Greece. In spite of the fact that both trends based on Christian tenets, the first one was more fantastic and far from the true picture of the world in compare the second one, which explained its structure using the ancient knowledge.

In the XV century in Ukraine were spread 'Cosmography' De-Sakrabosko (John Halifax) and 'Shestokryl' Immanuel bar-Jakob (Jewish scholar XIV century). In Ukrainian version of 'Shestokryl' is stated that the Earth had a shape of a ball [10, 11]. In the work also is said that the celestial bodies were moving owing to self –rotating of spheres. There are nine celestial circles. The Author of 'Cosmography' said: 'All sky are situated in each other. This structure is like to an onion'. In the same time 'The Earth in the centre of the Sky. It does not move.' [11, p. 178].

It is believed that Zachariah (Shariya) (scientist from Kyiv city) translated and disseminated 'Shestokryl' that was used for astronomical calculations. The appearance and spreading of 'Cosmography' and 'Shestokryl' in the second half of XV century can be considered as a new stage in the development of astronomical ideas in Ukraine. In the spirit of scientific traditions Renaissance, educated people had the opportunity to discover and to have a familiarization of Aristotle and Ptolemy's systems through these works.

The accumulation of physical and mathematical knowledge in Ukraine was favored by the activities of Ukrainian humanists of the period of the end of XV – the beginning of XVI. The ideas of Renaissance came into Ukraine owing to the education of Ukrainian people in the universities of Vienna, Padua, Bologna, Venice, Rome and Krakow. The period of creation of new generation of humanists was the end of XVI – the first half of XVII century. The representatives of humanists organized and took part in the life of cultural and educational centers. The most important among these centers was Ostroh (Ostrog) Academy, which combined the old Ukrainian and Greek-Byzantine traditions of educational achievements of Europe. There were taught 'seven free sciences', including mathematics, astronomy and philosophy. Its students studied the works on mathematics, astronomy, philosophy and physics, written mostly in Latin. Among them is 'Cosmography' by Bleu I., which contained information about the Copernicus's ideas about the structure of the world. 'Cosmography' by Bleu I. was translated into Russian language by Epiphany Slavinetsky (1645-1647) and was known under the titles 'Mirror of whole World…' and 'astronomical calendar of 1506' and others.

At the end of the first quarter of the XVII century, Kyiv had a status of the leading orthodox cultural and educational center. Its influence was strengthened by the formation of Bratsk School, printing activity of Kyiv Pechersk Lavra and the activities of the orthodox metropolitan department. At that time, in Kyiv was builded the Kyiv-Mohyla Academy. Its students formed an influential layer of educated clergy, who favored the spread of knowledge. The level of the courses studied in the Academy met the demands of West-European Higher Education. While studying theoretical courses, the main authorities were Aristotle and Thomas Aquinas [2].

At the end of XVII century Kyiv-Mohyla Collegium was achieved the status of the academy. It was introduced the course of theology. Training in Academy in high school lasted six years and involved two-year course of philosophy and four-year course of theology. Philosophy was divided into "natural philosophy" (followed by further in-depth studying of mathematics) and metaphysics. The natural philosophy included physics (cosmogony, meteorology and others) and physiological psychology. Metaphysics considered supernatural phenomena, their causes and general principles according to belief that 'the God creates the world' [12, 13].

Courses of philosophy that were read in Ukraine in the first half of XVIII century were similar to those taught at leading European universities. According to Ukrainian researchers, the lectures of Galileo in Padua University (1592 – 1610) were visited by 52 Ukrainian students. Many Ukrainian men's studied in Rome (among them are Joazaf Krokowski, Theophan Prokopovich, Teofilakt Lopatynskyi, Innokenty Gizel). In the Academy was formed a new philosophical worldview. Today, the researchers of philosophical heritage of Kyiv-Mohyla Academy believe that it had two directions of development of philosophy knowledge. One of them is a research and an education. Its representatives put more emphasis on the development of science, education, crafts, arts and education. It was so-called Peripatetic natural rationalistic line. It was supported by I.Gizel, J. Krokowski, Th. Prokopovich, G. Rodin, M. Kozachynska and others.

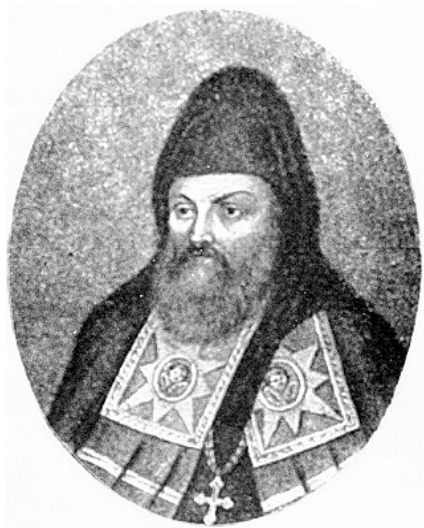

**3 Innokenty Gizel**

Innokenty Gizel (1600–1683) studied at the Mohyla collegium which later became known as the Kyiv–Mohyla Academy (National University of Kyiv-Mohyla Academy). He studied philosophy, theology, law and other sciences in the Zamojski Academy and at universities in Germany and England [14]. I.Gizel included knowledge about all directions of philosophy in his philosophy course 'The work of the whole philosophy' that taught at the Academy in the 1645-1647. Chronologically, his work was the first course of Nature Philosophy that was read at the Academy. Besides geocentric world system, I.Gizel studied the system of Copernicus. It was the first mention of the name of N. Copernicus in Ukraine in an educational school.

Professors of the Academy S.Kleshanski, S.Jaworski, I.Popovski were not satisfied with the Aristotle–Ptolemy's theory. So they tried to develop another theory of the world structure. They did not recognize Copernicus' theory correct, but during his teaching, they used Copernicus's tables and drawings. In this way, they gave their students possibility to decide themselves whether the theory was correct or not.

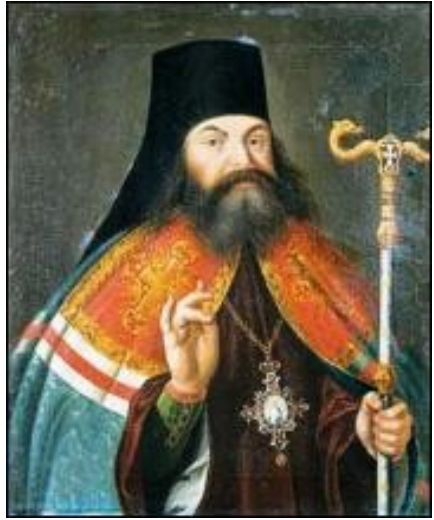

**4 Theophan Prokopovich**

Theophan Prokopovich (1677–1736) studied at Kyiv-Mohyla Academy and in the Academy in Rome. He listened to the lectures in the universities of Hale, Leipzig, Hyena and Kongsberg. In 1705–1716 he taught all the higher sciences at the Academy and was the first man who began to acquaint students with the teaching of Descartes, Locke and Bacon. He wrote about physics: "It 'fertilized' all art and considerable favored the life of human race" [15, p.115].

He defines the notion 'world' in 'Natural Philosophy or Physics' [2, p.113-502]. According to him, the world is the structure that consists of heaven, earth and other elements that are located between the heaven and earth. In other words, the world is 'the order and location of all that is saved God' [2, p.283]. Th. Prokopovich acquainted students with all common theories about the universe of that time. At the beginning, he taught the world system of Ptolemy. He mentioned that this theory was offered by Pythagoras. However, being an outstanding mathematician, Ptolemy explained the antique system in more detail [2, p.286]. Then, he taught Copernican's system. Th.Prokopovich emphasized that this theory was insufficient to explain the many complexities of astronomy.

He also said that the Earth does not move and the Sun moves. Then, he introduced the theory of Tycho Brahe.

Despite the fact that in his course Th. Prokopovich taught various systems of the world, but he believed that the world had been created by God. He mentioned that according to Holy Scripture, the world did not exist forever, 'Heaven and Earth were originally created' [2, p.296].

So, Ukraine scholars knew about all models of the universe that existed at that time. The schools gave students information about all existing at the time cosmological theories. After appearing of heliocentric system, Western Europe suffered the war between science and religion. However, Ukraine students were acquainted with not only with Ptolemy's system and Copernicus's theory, but also Descartes' and Kant–Laplace's one. It should be noted that until the end of the XVIIIth century a key role in explaining the origin and structure of the Universe had a religion and the Bible text.